\documentclass[fleqn,usenatbib]{mnras}
\usepackage{newtxtext,newtxmath}
\usepackage[T1]{fontenc}
\DeclareRobustCommand{\VAN}[3]{#2}
\let\VANthebibliography\thebibliography
\def\thebibliography{\DeclareRobustCommand{\VAN}[3]{##3}\VANthebibliography}

\usepackage{graphicx}	
\usepackage{amsmath}	
\usepackage{multirow}

\title{Leaky dust trap in the PDS 70 disk revealed by ALMA Band 9 observations}

\author[A. Sierra et al.]{
Anibal Sierra,$^{1}$
Myriam Benisty,$^{2}$
Paola Pinilla,$^{1}$
Laura Pérez,$^{3}$
Pietro Curone,$^{3}$
Kiyoaki Doi,$^{2}$
\newauthor
Stefano Facchini,$^{4}$
Daniele Fasano,$^{2,5}$
Sean Andrews,$^{6}$
Jaehan Bae,$^{7}$
John Carpenter,$^{8}$
Ian Czekala,$^{9}$
\newauthor
Andrea Isella,$^{10,11}$
Nicolas Kurtovic,$^{12}$
Francois Menard,$^{5}$
Richard Teague,$^{13}$
\\
$^{1}$Mullard Space Science Laboratory, University College London, Holmbury St Mary, Dorking, Surrey RH5 6NT, UK\\
$^{2}$Max-Planck Institute for Astronomy (MPIA), Königstuhl 17, 69117 Heidelberg, Germany\\
$^{3}$Departamento de Astronomía, Universidad de Chile, Camino El Observatorio 1515, Las Condes, Santiago, Chile\\
$^{4}$ Dipartimento di Fisica, Università degli Studi di Milano, Via Celoria 16, 20133 Milano, Italy\\
$^{5}$Université Côte d’Azur, Observatoire de la Côte d’Azur, CNRS, Laboratoire Lagrange, France\\
$^{6}$ Center for Astrophysics, Harvard \& Smithsonian, 60 Garden Street, Cambridge, MA 02138, USA\\
$^{7}$Department of Astronomy, University of Florida, Gainesville, FL 32611, USA\\
$^{8}$Joint ALMA Observatory, Avenida Alonso de Córdova 3107, Vitacura, Santiago, Chile\\
$^{9}$School of Physics \& Astronomy, University of St. Andrews, North Haugh, St. Andrews KY16 9SS, UK\\
$^{10}$Department of Physics and Astronomy, Rice University, 6100 Main St, Houston, TX 77005, USA\\
$^{11}$Rice Space Institute, Rice University, 6100 Main St, Houston, TX 77005, USA\\
$^{12}$Max-Planck Institut für Extraterrestrische Physik (MPE), Giessenbachstr. 1, 85748 Garching, Germany\\
$^{13}$Department of Earth, Atmospheric, and Planetary Sciences, Massachusetts Institute of Technology, Cambridge, MA 02139, USA\\
}

\date{}

\pubyear{\the\year{}}

\begin{document}
\label{firstpage}
\pagerange{\pageref{firstpage}--\pageref{lastpage}}
\maketitle

\begin{abstract}
We present new observations of the PDS\,70 disc obtained with the Atacama Large Millimeter/sub-millimeter Array (ALMA) in Band\,9 (671\,GHz) at 0.242$^{\prime\prime}$ resolution, which provide valuable insights into the spatial distribution of sub-millimetre grains in the disc. The data reveal a ring-like morphology, with a radial peak located between those previously observed at infrared wavelengths and longer millimetre observations. Additionally, we detect a tentative outer shoulder in Band 9 that is not observed at longer wavelengths. These findings suggest that small grains ($\sim 100 \mu$m) traced by Band\,9 may be escaping from the pressure bump both radially inwards and outwards, or may be tracing different disc layers than those probed at longer wavelengths.
A multi-wavelength analysis of the disc at millimetre wavelengths and the best fit to the spectral energy distribution shows the presence of centimetre grains around the ring location, where the dust surface density also peaks, compatible with dust trap models. The grain size in the disc cavity is not well constrained but is consistent with grains as small as 10\,$\mu$m, supporting the hypothesis that small dust grain filters through the cavity.
We use dust evolution models to demonstrate that a turbulent viscosity of $\alpha\,\gtrsim\,10^{-3}$ allows small grains to filter through the disc gap, while $\alpha\,\lesssim\,5\,\times\,10^{-3}$ is required to retain large grains in the pressure bump. The Band\,9 observations of PDS\,70 validate theoretical models and confirm the presence of pebble flux through the disc gap. 
\end{abstract}




\section{Introduction} \label{sec:introduction}
Dust grains in protoplanetary disks are the building blocks of planetesimals and the cores of planets. At early stages, collisions between sub-micrometre grains are the primary mechanism responsible for dust growth, from ISM-like grains to millimetre- or centimetre-sized grains \citep{Dominik_2007, Blum_2008}.
However, as dust grains grow, they decouple from the gas and lose angular momentum, causing them to drift toward the central star (radial drift) on timescales of a small fraction of a Myr, if nothing halts their inward migration \citep{Weidenschilling_1977}. This contradicts the presence of bright and dust-rich discs observed in dozens of protoplanetary systems with ages around 1 Myr \citep[e.g.,][]{Andrews_2018, Cieza_2019, Carpenter_2025}.

Several mechanisms, have been proposed to prevent the rapid radial migration of millimetre-sized grains \citep[e.g.,][]{Johansen_2009, Dzyurkevich_2010, Lesur_2014, Gonzalez_2017}, with dust trapping in pressure bumps being one of the most widely accepted theories \citep[e.g., ][]{Pinilla_2012b, Birnstiel_2013}. It is well known that the efficiency of dust trapping is grain size ($a$) dependent \citep[e.g., ][]{Takeuchi_2002, Pinilla_2012b, Flock_2015}. In particular, the radial drift as well as the efficiency of dust trapping depend on the interaction between dust and gas \citep{Whipple_1972, Weidenschilling_1977}, which is quantified by the Stokes number \citep[$St \propto a$,][]{Stokes_1851}. Dust grains with a Stokes number of 1 (corresponding to millimetre-sized grains under typical disc conditions) are effectively trapped in pressure bumps or rapidly drift toward the central star if no traps are present.

Dust grains are not only trapped by drag forces but also spread out due to diffusion. The level of dust diffusivity is controlled by the diffusion coefficient ($D$), which depends on the Stokes number as $D \propto (St^2 + 1)^{-1}$ \citep{Youdin_2007}. Therefore, small grains (with low Stokes numbers) are not effectively trapped in pressure bumps and are also more prone to diffusion.

The main observational signatures of traps in protoplanetary disks are rings and gaps (or deep cavities), which may be produced by embedded planets that create pressure bumps where dust grains become trapped \citep[e.g.,][]{Pinilla_2012b}.
The dust flux through the disc cavity, created by the presence of a planet opening a gap has been studied in 1D, 2D, and 3D simulations \citep[e.g.,][]{Rice_2006, Paardekooper_2006, Zhu_2012, deJuanOvelar_2013, Bae_2019, Drkazkowska_2019, Stammler_2023, Huang_2025}. They show that although the radial migration of large grains is halted, small grains filter through the cavity, feeding both the cavity and the inner region of the disc with dust. Moreover, collisions between large grains within the pressure bump continuously generate new small grains via fragmentation, supplying dust material to the gap and inner disc over several Myr for Jupiter-mass planets \citep{Stammler_2023}. This dust flux can increase near the planet location, where the planet gap is not axisymmetric \citep{Drkazkowska_2019}.

From an observational perspective, observations at a wavelength $\lambda$ are mainly sensitive to grain sizes of $a \sim \lambda / \pi$ \citep[e.g., ][]{Draine_2006, Ricci_2010, Birnstiel_2018}. 
Therefore, dust segregation by drag forces and diffusion has important consequences on the morphology of the discs observed at different wavelengths \citep{Zhu_2012}, and the radial brightness distribution may significantly vary with wavelength.
In particular, observations at shorter wavelengths are expected to be more sensitive to small grain sizes that are filtered through disc cavity, while longer wavelengths will trace larger grains that thought to be trapped in pressure bumps.

The disc around PDS 70, where two forming planets have been confirmed using several techniques and multi-wavelength observations \citep{Keppler_2018, Muller_2018, Mesa_2019, Benisty_2021}, is an excellent laboratory to test these models and look for observational evidence that would suggest filtering of small grain sizes through the deep cavity created by the two forming planets. 
The estimated mass of the two planets are $3.2^{+3.3}_{-1.6} M_{\rm Jup}$ and $7.5^{+4.7}_{-4.2} M_{\rm Jup}$ for PDS 70b and PDS 70c, respectively \citep{Wang_2021}. These masses are sufficient to open a deep disc cavity and prevent the flow of dust toward the inner disc.

However, millimetre observations and SED constraints at infrared wavelengths of PDS 70 have revealed the presence of an inner disc \citep{Dong_2012, LongZachary_2018, Benisty_2021, Doi_2024}, with a size of $\sim 18$\,au (a factor of 4.2 times smaller than the disc cavity), which has been explained by the filtration of very small grain sizes through the disc gap, continuously replenishing the inner region of PDS 70 with dust material and accounting for its long-term survival \citep{Bae_2019, Pinilla_2024}. This hypothesis has not been tested to date, as only Band 7 and Band 3 millimetre data have been published so far.
Evidence of dust filtration has also been suggested by the observation of infrared peaks interior to the millimetre peaks \citep{Zhou_2025}, and by the presence of water vapour in the inner disc \citep{Perotti_2023}, which can be transported through the gap by ice-coated small dust grains.

In this work, we report sub-mm observations of the disc around PDS 70 at ALMA Band 9 (671\,GHz, 447\,$\mu$m), where the dust continuum ring is resolved (Section \ref{sec:observations}), and compared with other ALMA observations at longer wavelengths (Section \ref{sec:results}) in the image plane and from visibilities. We found evidence for small grain filtration in the disc gap, which is discussed in Section \ref{sec:DustModels}. Our key findings are discussed in Section \ref{sec:discussion}, and present the conclusions in Section \ref{sec:conclusions}.

\section{Observations}\label{sec:observations}
The disc around PDS 70 was observed by ALMA in Band 9 (671\,GHz) in April 2022 (Project code: 2021.1.01058.S, PI: Myriam Benisty) using the configuration C43-3. The spectral setup of the observation was configured with eight spectral windows, covering frequencies between 659.01–660.89 GHz, 660.95–662.82 GHz, 662.91–664.78 GHz, 664.91–666.78 GHz, 675.08–676.95 GHz, 677.08–678.95 GHz, 679.03–680.91 GHz, and 680.97–682.85 GHz, at a spectral resolution of 1128.91 kHz and a total bandwidth per spectral window of 1.875 GHz. The disc was observed for a total on-source time of 36.45 minutes, using 45 antennas with baselines ranging from 15 to 500\,m.

We detected five lines centred at $\sim$661.46, 665.67, 666.63, 676.06, and 676.19 GHz. These values do not coincide with the rest frequencies of known molecular transitions and cannot be explained by Doppler shift at the systemic velocity of PDS 70 \citep[$5505 \pm 2$ m s$^{-1}$,][]{Keppler_2018}. After imaging the line cube, we double-checked that these lines are not part of the disc but appear as unstructured emission across the field of view, consistent with residual atmospheric features.

The continuum-only dataset was obtained by looking and flagging line emission in each spectral window and averaging the output data to a maximum channel width of 125 MHz. 
The continuum-only data was self-calibrated using CASA version 6.4.2.27 \citep{CASA_2022}. We used the task \texttt{tclean} \citep{Cornwell_2008} with a Briggs robust parameter of 0.5 to image the dataset after each self-calibration iteration, a multi-scale algorithm with angular scales corresponding to point sources, one beam size, and two beam sizes, and a clean threshold of $2\sigma$. During each self-calibration iteration, we used the mode \texttt{applymode=`calonly'} within the task \texttt{gaincal} to calibrate the data only and did not apply flags from the solutions.
After three rounds of self-calibration (with time intervals equal to \texttt{solint= `inf', `360s', `120s'}), where we combined spectral windows and scans, the signal-to-noise ratio (SNR) increased by 178\%, 38\%, and 10\% with respect to the previous iteration, respectively. No further phase self-calibration iterations were performed, as the SNR did not significantly improve for smaller time intervals. A final amplitude self-calibration iteration was applied using \texttt{solint= `inf'} while combining spectral windows and scans, leading to a 4\% improvement in SNR. The final dust continuum map was imaged by cleaning down to $1\sigma$, using a Briggs robust parameter of 0.0, and a Gaussian uvtaper of 0.10$^{\prime \prime}$ $\times$ 0.22$^{\prime \prime}$; 63\,deg (to circularize the beam), resulting in a restored beam size of 0.241$^{\prime \prime}$ $\times$ 0.238$^{\prime \prime}$; -6.1\,deg. A final convolution was applied using the task \texttt{imsmooth} to get a circular beam of 0.242$^{\prime \prime}$, and an rms of 0.37 mJy beam$^{-1}$.

In addition to the Band 9 observations, in this work we use and compare the Band 7 (350 GHz), Band 6 (220 and 260 GHz), and Band 3 (100 GHz) observations of PDS 70 from \cite{Benisty_2021, Facchini_2021, Doi_2024, Law_2024, Rampinelli_2024}. We use the self-calibrated measurement sets from those works and image them using robust values of 1.0, 0.0, and 2.0, respectively. These values provide a good balance between sensitivity and angular resolution.

\section{Results}\label{sec:results}

\subsection{Dust continuum Band 9 Map}
\label{sec:B9_data}

\begin{figure*}
    \centering
    \includegraphics[width=\linewidth]{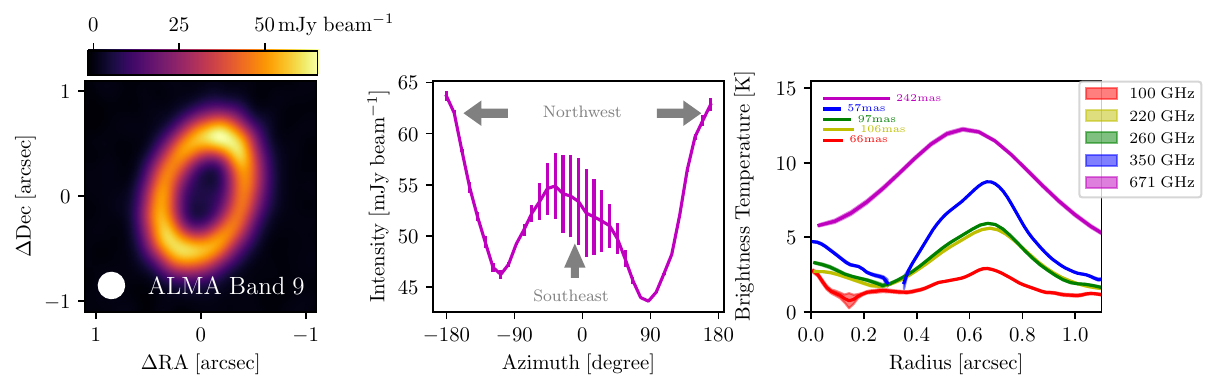}
    \caption{Left: Dust continuum image at ALMA Band 9 (671 GHz). 
    Middle: Azimuthal profile around the ring. The arrows indicate the azimuth of the northwest and southeast major axes.
    Right: Brightness temperature radial profile of PDS 70 extracted from the ALMA images at different wavelengths: 100 GHz (red), 220 GHz (yellow), 260 GHz (green), 350 GHz (blue), 671 GHz (purple). The horizontal lines show the angular resolution for each data set.}
    \label{fig:continuum-map}
\end{figure*}

The left panel of Figure \ref{fig:continuum-map} shows the dust continuum map at ALMA Band 9, where a ring-like structure is observed, as in previously resolved images at millimetre and infrared observations \citep{Keppler_2018, Muller_2018, Benisty_2021, Doi_2024}. The map shows two bright azimuthal asymmetries along the disc's major axes, a property that may be associated to an optically thin and geometrically thick disc \citep{Doi_2021}. However, as discussed in Section \ref{sec:OptDepth}, the angular resolution makes it difficult to draw conclusions about the optical depth regime in Band 9 from the map itself.

The middle panel shows the azimuthal profile centred at 0.6$^{\prime \prime}$ (radial position of the peak) and within a radial extent of one beam size. This profile (as well as all radial profiles in the image plane in the following Figures) is computed using \textsc{GoFish} \citep{Teague_2019} adopting inc = 51.7 deg, PA = 160.4 deg \citep{Benisty_2021}. This geometry is assumed throughout the following sections.
The emission in the northwest is higher than that in the southeast by a factor of $\sim 1.2$. This asymmetry coincides with that previously observed at higher angular resolution in the dust continuum observations from Band 7 to Band 3.

The right panel show the radial profile of the brightness temperature (purple line), together to the brightness temperature profile at 350 GHz, 260 GHz, 220 GHz, and 100 GHz. We do not assume the Rayleigh-Jeans approximation when computing the brightness temperature, as this approximation can introduce significant errors at high frequencies or low disc temperatures.

The peak position in the Band 9 data is shifted inward compared to the other wavelengths. To avoid any possible angular resolution effect, we confirm this offset by modelling the dust continuum visibilities in Section \ref{sec:visibilities} and discuss its possible origin in Section \ref{subsec:Origin_Peak}.
At the resolution of Figure \ref{fig:continuum-map}, the ring appears symmetric with respect to the peak, and the inner disc (as that observed at longer wavelengths) is not detected. 

\subsection{Visibility modelling}
\label{sec:visibilities}

\begin{figure*}
    \centering
    \includegraphics[width=0.9\linewidth]{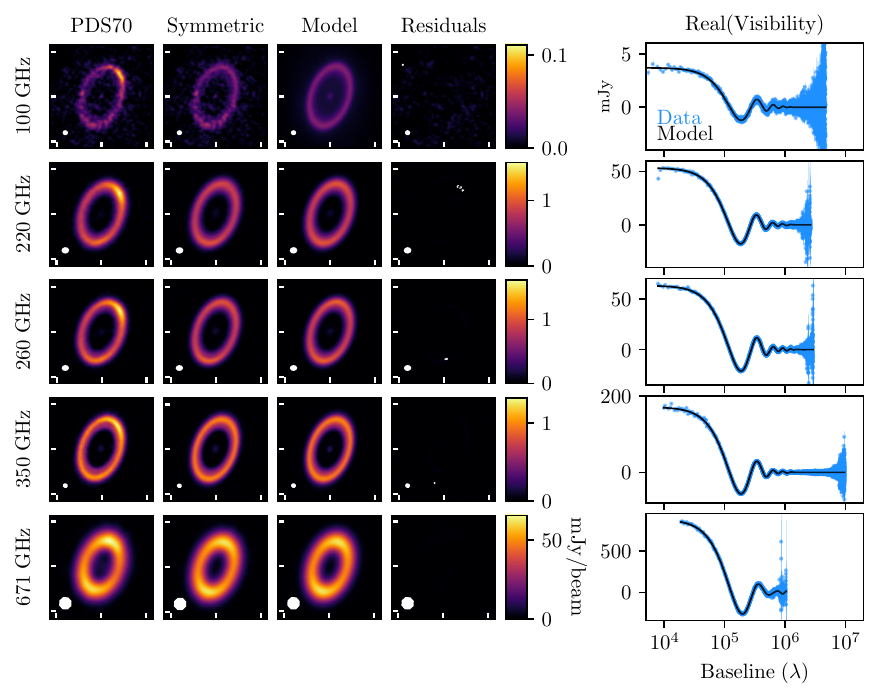}
    \caption{Visibility modelling of the disc around PDS 70 at different wavelengths.
    Left: tclean image of the dust continuum data. Middle left: tclean image of the symmetric dust continuum data. Middle right: tclean image of the Frankenstein model. Right: tclean map of the residual visibilities (Symmetric - Model). The white iso-contour show the region with residuals above/below $\pm 5\sigma$.    
    The units of all colour bars are mJy beam$^{-1}$.
    Right panels show the real part of the continuum visibilities (blue dots) as a function of the uv-distance (baseline). The flux units in all right panels are mJy.
    The solid black line shows the visibility Frankenstein model.
    }
    \label{fig:visibility_modelling}
\end{figure*}

We conducted visibility modelling of the dust continuum emission of the disc around PDS 70 using \textsc{Frankenstein} \citep{Jennings_2020}, a Python package that fits the radial brightness profile of interferometric data using the real part of the visibilities. \textsc{Frankenstein} assumes axisymmetric disc morphologies. 
However, the disc around PDS 70 exhibits an asymmetric arc-like feature in the northwest (dust continuum maps in the left column of Figure \ref{fig:visibility_modelling}). Therefore, before using \textsc{Frankenstein}, we compute the visibilities of the azimuthal asymmetry following a methodology similar to that in \cite{Andrews_2021}, and applying the arc mask described in Appendix B of \cite{Benisty_2021}. 

First, we estimate the level of emission in the ring by computing the azimuthally-average intensity radial profile using the clean components outside the mask. We then subtract this emission within the arc mask and use the \texttt{sampleImage} task in \textsc{galario} \citep{Tazzari_2018} to compute its visibilities. Finally, these visibilities are subtracted from the full data sets.
This methodology is applied in all wavelengths, including the Band 9 data, where the azimuthal asymmetry is not as evident as in the other datasets. The resulting visibilities (asymmetry subtracted datasets) are imaged and presented in the middle-left column of Figure \ref{fig:visibility_modelling}.

We then perform \textsc{Frankenstein} fits using the visibilities from the asymmetry subtracted datasets, fixing the fit hyper-parameters to the upper values recommended by \cite{Jennings_2020}, $\alpha_{\rm f} = 1.3$, $w_{\rm s} = 10^{-1}$ at all wavelengths, except in Band 9, where we used $\alpha_{\rm f} = 1.3$, $w_{\rm s} = 10^{-4}$. This choice is based on the visibility coverage of the different observations. While the Band 3, 6, and 7 data have maximum baselines of a few M$\lambda$, the Band 9 data are only marginally close to 1 M$\lambda$ (right panel in Figure \ref{fig:visibility_modelling}). Therefore, a lower $w_{\rm s}$ value is chosen for Band 9 to achieve a better fit to the power spectrum and to be sensitive to visibilities from the longest baselines \citep[see discussion on the hyper-parameters in][]{Jennings_2020, Jennings_2022}. The effect of $\alpha_{\rm f}$, $w_{\rm s}$ on the reconstructed intensity radial profile in Band 9 is explored in Appendix \ref{app:VisFit}.

The disc centres are computed by minimising the imaginary part of the visibilities of the asymmetry subtracted data sets \citep[e.g.,][]{Isella_2019}. We obtained offsets of ($\Delta$RA, $\Delta$Dec) = (0.05 mas, -0.01 mas), (-1.7 mas, -1.5 mas), (1.2 mas, -0.5 mas), (2.5 mas, -1.3 mas), (-15.0 mas, -14.7 mas) for Band 3, 6 (at 220 GHz), 6 (at 260 GHz), 7, and 9, respectively. The inclination and position angle are fixed to the same values used to compute the radial profiles in the image plane (Section \ref{sec:observations}). The \textsc{Frankenstein} fits are shown as a black solid line in the right panels of Figure \ref{fig:visibility_modelling}.

The model visibilities are also imaged in \textsc{CASA}, and they are shown in the middle-right panel of Figure \ref{fig:visibility_modelling}. The right panels show the maps of the residuals (symmetric data - model), where negligible residuals structures above $5\sigma$ or below $-5\sigma$ are inferred.
The residual map at Band 9 is also shown in Appendix \ref{app:B9_residuals} in units of the rms.
The intrinsic resolution of the visibility modelling is estimated following the methodology in Appendix A of \cite{Sierra_2024b}, where the spread of a delta ring is measured given the uv-coverage and SNR of each observation. The resulting resolutions are 55.8, 57.0, 51.1, 19.1, and 176.5 mas for Band 3, 6 (at 220 GHz), 6 (at 260 GHz), 7, and 9, respectively. Compared to the synthesized beam in \textsc{CASA}, these effective resolution elements are smaller by factors of 1.2, 1.8, 1.9, 3.0, and 1.4, respectively.

\begin{figure*}
    \centering
    \includegraphics[width=0.8\linewidth]{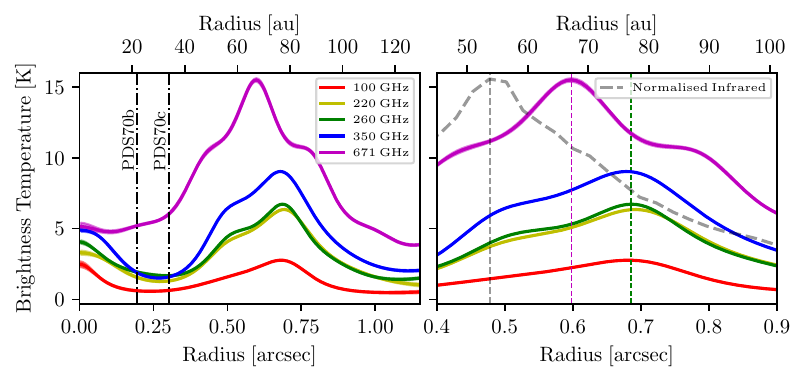}
    \caption{Left: Radial profiles of the brightness temperature obtained from the visibility fit at different ALMA wavelengths (see colour code). The vertical dashed-dotted lines indicate the orbits of PDS 70b, and PDS 70c. 
    Right: Zoom-in around the ring peaks. The grey dashed line shows the normalised brightness radial profile from infrared observations at H band ($\lambda = 1.625 \mu$m). The radial positions of the peaks are shown as vertical dashed lines.}
    \label{fig:BrightnessTemperature}
\end{figure*}

The brightness temperature radial profile obtained from the visibility modelling is shown in the left panel of Figure \ref{fig:BrightnessTemperature} for all the millimetre wavelengths. In addition, the right panel shows a zoom around the peaks and the normalised brightness radial profile from infrared observations obtained with SPHERE at H band \citep[$\lambda = 1.625 \mu$m, ][]{Wahhaj2024} for comparison.
The Band 9 observations present a different morphology compared with millimetre observations at longer wavelengths. In particular, the radial position of the peak has an inward offset, as already observed in the radial profiles from the image plane (Figure \ref{fig:continuum-map}).
The position of the peak in Band 9 is $\sim$ 0.60$^{\prime \prime}$ (67.4 au), while for the rest of the wavelengths, it is located at $\sim 0.68 ^{\prime \prime}$  (76.4 au), resulting in a 9 au offset. The position of the infrared peak is $\sim$ 0.48$^{\prime \prime}$ (54.0 au), resulting in an inward offset of 13.4 au with respect to the Band 9 peak. The displacement of the peak with wavelength is studied in terms of dust evolution segregation models in Section \ref{sec:DustModels}.

The position of the inner shoulder at Band 9 is also radially offset inwards, compared with Band 7, 6, and 3. Additionally, a second outer shoulder, located beyond the peak position of the longer wavelengths, is present. The inner and outer shoulders are only inferred from the visibility modelling when the fit to the power spectrum of the data is performed using a \textsc{Frankenstein} hyper-parameter of $w_{s} \lesssim 10^{-3}$ (Appendix \ref{app:VisFit}). Therefore, they should be interpreted with caution. However, the peak position is a robust measurement in all cases. 

\subsection{Multi-wavelength analysis}
\label{sec:Multiwave}

\begin{figure}
    \centering
    \includegraphics[width=\linewidth]{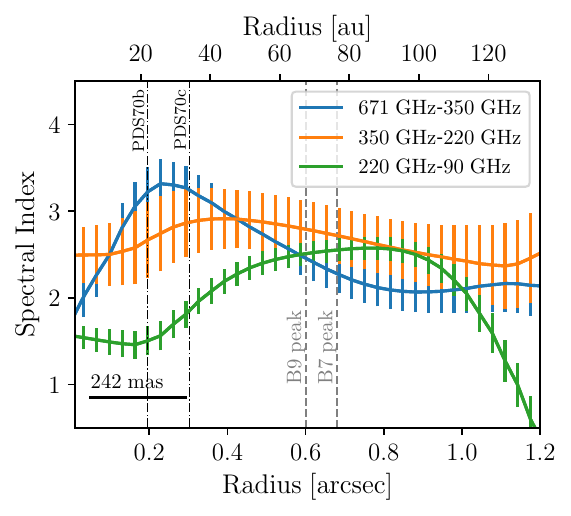}
    \caption{Spectral index radial profile of the disc around PDS 70 at an angular resolution of 242$^{\prime\prime}$ (indicated in the left bottom as a solid line), and between different millimetre bands (see colour code). The vertical dashed lines show the position of the peak at Band 9 and Band 7, while the vertical dashed-dotted lines indicate the orbit of PDS 70b and PDS 70c.}
    \label{fig:SPI}
\end{figure}

Figure \ref{fig:SPI} shows the spectral index radial profile between different bands computed from the intensity profiles in the image plane at a common angular resolution of 0.242$^{\prime\prime}$. The central emission of the disc (around one beam size) is contaminated by free-free emission at Band 3 \citep{Doi_2024}, presenting spectral indices below 2. The spectral index between Band 9 and Band 7 reaches a maximum around the orbits of PDS 70b and PDS 70c, where the disc has a deep gap. In contrast, it tends to be lower than the spectral index computed between Band 7 and Band 6 around the disc ring.

Intensity radial profiles at several wavelengths (such as those in Figure \ref{fig:continuum-map} or \ref{fig:BrightnessTemperature}) can be used to infer physical disc properties such as dust temperature ($T_{\rm d}$), surface density ($\Sigma_{\rm d}$), and maximum grain size ($a_{\rm max}$) by modelling the spectral energy distribution and taking into account the effects of scattering \citep[e.g.,][]{Carrasco-Gonzalez_2019, Macias_2021, Sierra_2021, Sierra_2024b}.
The methodology in these previous works can be applied to multi-wavelength data at the same angular resolution. However, as noted in previous sections, the resolution of the Band 9 observations is lower than that of the rest of the data, limiting the inference of dust properties at high angular resolution.
However, we followed the approach proposed by \cite{Viscardi_2025}, who studied how high and low angular resolution can be simultaneously implemented in this kind of multi-wavelength analysis. They propose the implementation of two multi-wavelength fits: the first includes all the observations at low angular resolution, while the second includes only the high angular resolution observations, using the results of the former as prior distributions.

\begin{figure*}
    \centering
    \includegraphics[width=\linewidth]{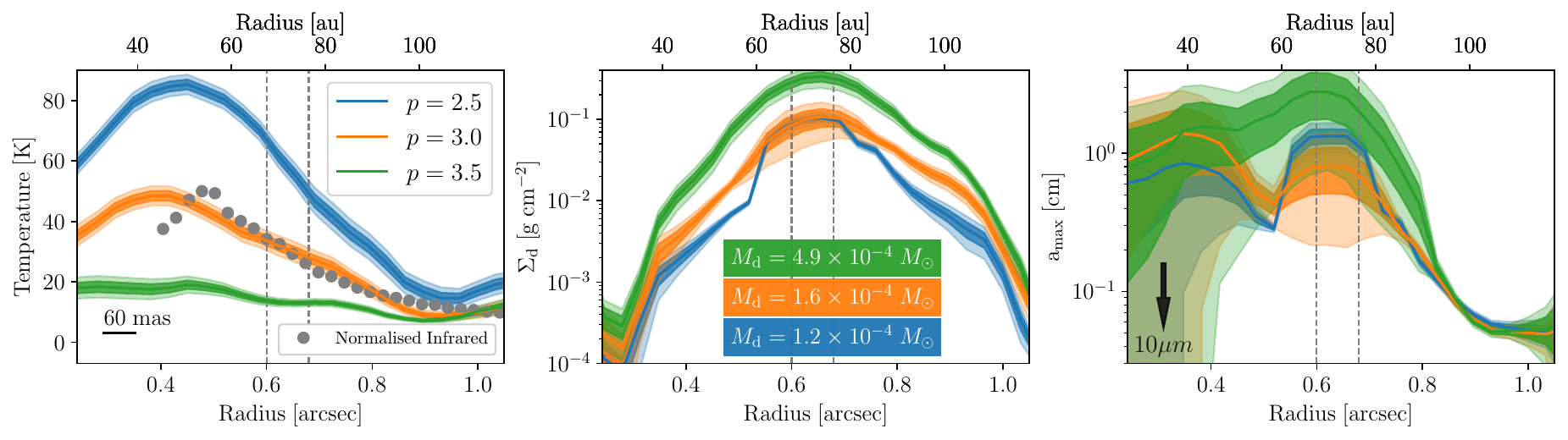}
    \caption{Dust properties constraints inferred from multi-wavelength analysis for different slopes of the particle size distribution $p$}. Left: Dust temperature. Middle: Dust surface density. Right: Maximum grain size. 
    The angular resolution is indicated in the bottom left corner of the left panel. The shaded areas show 1,2 $\sigma$ uncertainties, and the solid line is the best-fit value.
    The dots in the left panel shows the normalised brightness radial profile from infrared observations at H band.
    The total dust mass is shown in the bottom of the middle panel for each model.
    The arrow in the right panel indicates that the error bar in the inner disc extends up to $10 \mu$m.
    The dashed vertical lines show the position of the peak at Band 9 and Band 7.
    \label{fig:Multiwave}
\end{figure*}

Therefore, we use the methodology in Section 4 of \cite{Sierra_2024b} to perform both fits. In the first one, the priors follow a log-uniform distribution. The explored parameter space is $ 5 < T_{\rm d}[\rm K] < 300$, $-6 < \log_{10} (\Sigma_{\rm d} [\rm g \ cm^{-2}]) < 3$, $ -3 < \log_{10} (a_{\rm max} [\rm cm]) < 1$. We use the DSHARP opacities \citep{Birnstiel_2018}, and assumed that the grain size distribution follows $n(a)da \propto a^{-p}da$, with grain radii between $a_{\rm min} = 0.5 \mu$m and $a_{\rm max}$, where the latter is a free parameter of the fit.
We used three values for the particle size distribution: $p = 2.5, 3.0, 3.5$. The former and the latter represent the cases when the particle size distribution is dominated by drift and fragmentation, respectively \citep{Birnstiel_2012b}.

The Chi-squared is defined as 
\begin{equation}
    \label{Eq:Chi2}
    \chi^2 = \sum_{n} w_{\nu_n} \left( \frac{I_{\nu_n} - I_{\nu_n}^{\rm model}}{\Delta I_{\nu_n}} \right)^2,
\end{equation}
where $I_{\nu_n}^{\rm model}$ is the model intensity predicted from the free parameters using equation (9)-(10) in \cite{Sierra_2020}, $I_{\nu_n}, \Delta I_{\nu_n}$ are the observed intensities and their uncertainties, and $w_{\nu_n}$ are the relative weights of each band. The weights are fixed at 1, except for Band 9, which is assigned a value of 2 in the first fit to provide a stronger constraint on the disc temperature traced by the optically thicker wavelength. This procedure is applied independently to each radial bin, using radial intensity profiles at 671, 350, 220, and 90 GHz, computed from images convolved with a common circular beam of 0.242$^{\prime \prime}$.

For the second fit, we use the intensity radial profiles at 350, 220, and 90 GHz obtained from the Frankenstein fits and convolved at a common angular resolution of 0.06$^{\prime \prime}$. Band 9 is no longer used in the second fit due to its relatively low angular resolution compared with the rest of the data. The explored parameter space is the same as in the first fit, and the chi-squared is defined as in Equation (\ref{Eq:Chi2}), with weights equal to 1 at all wavelengths. 
Since the results from the first fit include Band 9, which is the most sensitive tracer of dust temperature within our multi-wavelength coverage, the priors for the dust surface density and maximum grain size are set to be log-uniform, and a temperature prior is introduced, given by
\begin{equation}
    \label{eq:Prior}
    p(T_{\rm d}) = \exp \left[ -\frac{1}{2} \left( \frac{T_{\rm prior} - T_{\rm d}}{\sigma_T} \right)^2 \right],
\end{equation}
where $T_{\rm prior}$, and $\sigma_T$ are the best fit model and its uncertainty, obtained from the first fit.

The results of the multi-wavelength fit at high angular resolution are shown in Figure \ref{fig:Multiwave}, and we show the results from the low-angular resolution fit in the case of $p=3.0$ in Appendix \ref{app:DustProperties}.
We have discarded the fit within 0.242$^{\prime\prime}$ to avoid filling factor issues when computing the intensity radial profiles in the image plane \citep{Thompson_2017}, and possible artifacts in the inner disk structure when modelling the visibilities \citep[see discussion in][]{Jennings_2020, Jennings_2022}.

In the three models, the best-fit dust temperature tends to decrease with increasing radius, and the dust surface density peaks around the radial location of the brightness peaks observed in Bands 9 and 7, with an outer shoulder that spatially coincide with the shoulder observed in Band 9, and where the dust temperature has a local minimum.

We include the normalized radial profiles of the infrared observations of PDS 70 in the H band ($\lambda = 1.625\ \mu$m) for reference in the left panel, as these short wavelengths primarily trace scattered light from the disc surface. In a flaring disc, this surface geometry may be sensitive to relative temperature gradients, as it is directly exposed to the stellar irradiation from the central star.
The slope of the infrared profile between $0.5^{\prime \prime}$ and $0.9^{\prime \prime}$ is $-2.38 \pm 0.06$, while for the dust temperatures are $-2.86 \pm 0.27, -2.60 \pm 0.36$ and $ -1.41 \pm 0.22$ for $p=2.5, 3.0$ and $3.5$, respectively. Since $p = 3.0$ yields the closest match, we normalized the infrared profile to this case.
For a flared disc with a scale height $h \propto r^\beta$, the infrared surface brightness is expected to scale as $r^{\beta -3}$ \citep{Whitney_1992}. Therefore, for PDS 70 we obtain $\beta = 0.61 \pm 0.06$, indicating that the aspect ratio decreases with radius $h/r \propto r^{-0.39 \pm 0.06}$.

The best fit for the maximum grain size tends to decrease with radius, from a $\sim$ 1 cm in the ring to $\sim 500 \mu$m beyond 100 au. The error bar in the inner disc and within the gap is significant, making possible the presence of small grain sizes ($a_{\rm max} < 100 \mu$m).
The best-fit for the grain size and dust surface density peaks around the ring and are consistent with dust trap signatures \citep[e.g.,][]{Pinilla_2012}, although within the uncertainties they are also consistent with a monotonically decreasing function.
The estimated dust surface density (and therefore the total dust mass) depends on the value of $p$. This dependence arises from differences in the dust opacity properties, as explained below: a larger maximum grain size ($a_{\rm max}$) is required to reproduce the same spectral index for increasing $p$ (as shown by our fits and consistent with Figure 4 in \cite{Birnstiel_2018}). Since the opacity coefficient decreases with increasing $a_{\rm max}$, a higher column density is needed to produce the same observed intensities.

The inferred dust mass is also expect to vary with resolution. For example, in the case of $p=3.0$, it lies between $1.6\,\times\,10^{-4} \ M_{\odot}$ ($0.17 M_{\rm J}$) when computed at 0.06$^{\prime\prime}$ resolution, and $4.2\,\times\,10^{-4}\,M_{\odot} (0.44\,M_{\rm J}$) when computed at 0.242$^{\prime\prime}$ resolution (Appendix \ref{app:DustProperties}). These total masses may be overestimated by a factor of 2 or 3 if no fully optically thin observations are included in the analysis \citep{Viscardi_2025}, but the values computed in this work
are comparable to the total dust mass estimated in \cite{Benisty_2021} when assuming a radially constant grain size of 1 mm or 1 cm.

\begin{figure}
    \centering
    \includegraphics[width=\linewidth]{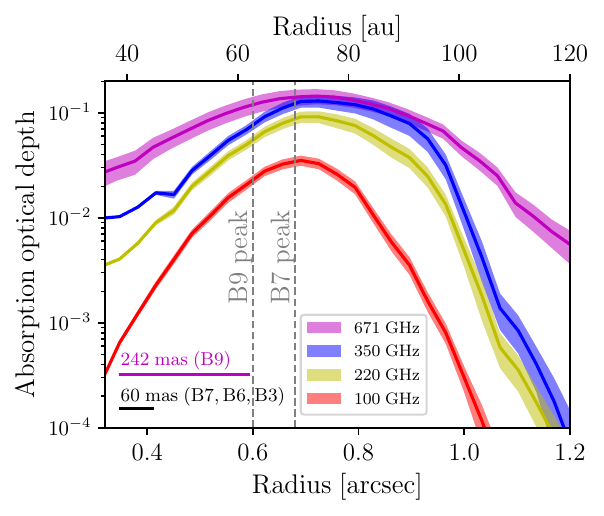}
    \caption{Optical depth radial profiles from the dust continuum multi-wavelength modelling (see colour code). The angular resolution for each band is indicated in the bottom-left corner. The dashed vertical lines show the position of the peak at Band 9 and Band 7.} 
    \label{fig:Opt-Depth}
\end{figure}

The absorption optical depth ($\tau_{\kappa_{\nu}} = \Sigma_{\rm d} \kappa_{\nu}$) computed from the multi-wavelength analysis are shown in Figure \ref{fig:Opt-Depth} for the case of $p=3.0$. We obtained the same optical depths for the models with $p = 2.5$ and $p = 3.5$. This occurs because, although higher dust surface densities are derived for $p = 3.5$, the lower opacity of the larger grain sizes in that model compensates for the increased dust mass. As a result, the product $\tau_{\nu} = \Sigma_{\rm d} \kappa_{\nu}(a_{\rm max})$ remains unchanged.

The profiles at Band 7, 6 and 3 are shown at 0.06$^{\prime\prime}$ resolution, while for Band 9 we present the profile at 0.242$^{\prime\prime}$ resolution. The optical depth regime of the data is discussed in Section \ref{sec:OptDepth}.

\section{Leakiness of dust particles in PDS 70 gap}
\label{sec:DustModels}

\begin{figure*}
    \centering
    \includegraphics[width=\linewidth]{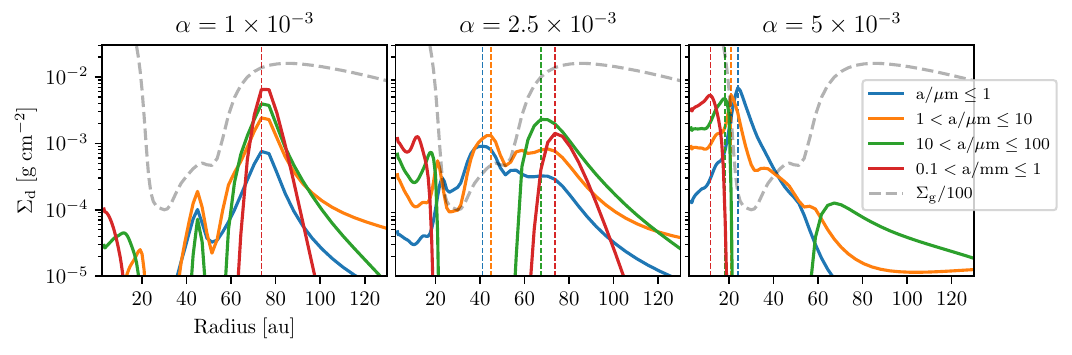}
    \caption{Dust segregation models for the disc around PDS 70. Different coloured lines (see colour code) show the dust surface density distribution for different grain sizes. The turbulent viscosity parameters in each simulation are $\alpha = 1\times 10^{-3}, 2.5\times 10^{-3}, 5 \times 10^{-3}$ from left to right, respectively. The dashed grey line is the gas surface density scaled by a factor of 1/100.
    }
    \label{fig:Segregation_Models}
\end{figure*}

PDS\,70 is an excellent source to study evolution of dust in the presence of giant planets in the disc, in particular to understand how much dust material can flow through the gap from the outer to the inner disc despite the deep gap. When a planet or multiple (giant-) planets interact with their parental disc, they can open a gap, which can lead to dust trapping in the outer edge of the gap \citep[e.g., ][]{Pinilla_2016, Drkazkowska_2019}. The trapping of dust particles depends on the shape of the gap that determines the pressure gradient and hence how much  the radial drift of particles is reduced at the gap outer edge. The trapping also depends on how well the particles are coupled to the gas and on the level of dust diffusion, as higher diffusion helps particles to escape the pressure bump. 

\cite{Pinilla_2024} performed tailored models of dust evolution for PDS\,70 to understand how the inner disc can survive for million-year timescales despite the presence of a deep gap carved by the massive planets in the system. They demonstrated that only very small particles ($<0.1\,\mu$m) can be  diffused and dragged along with the gas throughout the gap opened by the planets. Their models can explain the inner disc emission that has been revealed by ALMA in PDS\,70 and the pebble fluxes needed to explain the water abundances inferred from JWST observations \citep{Perotti_2023}. 

Nonetheless, their models do not predict the radial segregation that we find between the Band 9 and the rest of the frequencies from the ALMA observations. In \cite{Pinilla_2024}, they predict that all of the intensity profiles from the ALMA Bands analysed in their work would peak at the same location (see their Fig.~5). One possible explanation  to solve this discrepancy is that dust diffusion is higher, allowing small particles to easily diffuse out of the gap. As a consequence, a potential radial segregation can exist in the emission observed at different wavelengths. To test this idea, we perform two new dust evolution models based on the best models presented in \cite{Pinilla_2024}, with two different values of the disc viscosity that controls the dust radial diffusion. These models assumed a gap carved by a 2.5 and 5.0\,$M_{\rm{Jup}}$ from the models done in \cite{Bae_2019}.  The simulations assumed initial grain sizes from 0.01-1\,$\mu$m that are allowed to grow or fragment due to collisions. The fragmentation velocity is taken to be 10 m\,s$^{-1}$. In the models by \cite{Pinilla_2024}, the disc viscosity is assumed $\alpha=10^{-3}$, and we only change from these models the values of $\alpha$ to be either $\alpha=2.5\times10^{-3}$ or $\alpha=5.0\times10^{-3}$. The grain size fragmentation limits due to turbulence in the ring \citep[where the gas surface density is $\Sigma_{\rm g} = 1.47$ g cm$^{-2}$, ][]{Pinilla_2024} are $a_{\rm frag} = 130.1, 52.0$, and 26.0 $\mu$m for increasing values of $\alpha$, respectively.

Figure\,~\ref{fig:Segregation_Models} shows the dust density distribution from the dust evolution model after 1\,Myr of evolution for three different ranges of grains sizes and for three different values of disc viscosity. The left panel of Fig.~\ref{fig:Segregation_Models} shows the results for $\alpha=10^{-3}$ presented in \cite{Pinilla_2024}, where the dust surface density peaks at the same location independent of the grain size, including the very small grains ($a<1\mu$m). This is a consequence of efficient dust trapping in the pressure maxima and the enhancement of the small grains in this location is due to the efficient fragmentation of the pebbles into small dust particles. In this case, some of the small particles move through the gap, grow and replenish the inner disk of dust particles. 

The middle panel of Fig.~\ref{fig:Segregation_Models}  shows the case of $\alpha=2.5\times10^{-3}$. The increase of the disc viscosity has two main effects in the simulations. First, the turbulent velocities of particles increase, which makes the maximum grain size set by the fragmentation of particles to decrease ($a_{\rm frag} = 52$ $\mu$m in the ring), especially within the pressure bump where velocities due to drift are reduced.
The second effect is that the dust diffusion is higher, allowing particles to move along with the gas and elude the trapping, especially for the small particles. The first effect enhances the second as dust fragmentation reduce the possibility that particles can grow big enough to avoid being diffused out of the pressure bump.  As a result, the dust surface density of the large particles  ($a>10\mu$m) peaks at the location of pressure maxima, but for the smaller grains, the dust surface density peaks closer-in. The amount of dust that crosses the gap is much more than in the case of $\alpha=10^{-3}$, which makes the dust surface densities of different grain sizes to increase. 

From the right panel of Fig.~\ref{fig:Segregation_Models} we can see that with $\alpha=5.0\times10^{-3}$ trapping becomes inefficient due to the decrease of the grain size fragmentation limit ($a_{\rm frag} = 26 \ \mu$m in the ring)  and the high dust diffusion. As a result,  there is not a predominant ring-like structure of the large grains. Comparing with the results from the ALMA observations, we suggest that an increase of dust diffusion near the outer edge of the gap can lead to a higher leakiness of dust particles in the gap. As a consequence, there is a radial segregation of the dust density distribution among different grain sizes that can potentially explain the shift observed in Band 9 with respect to rest of the Bands. However, this increase in diffusion cannot be very high as it would prevent dust trapping and the formation of a bright ring as observed with ALMA in different transition discs \citep{deJuanOvelar_2016}.  From the simulations in this work, the intermediate case with $\alpha = 2.5 \times 10^{-3}$ is the best one explaining the observations, even though reproducing in detail the fluxes and intensity radial profiles is beyond the scope of this work.

\section{Discussion}\label{sec:discussion}

\subsection{Origin of the peak shift in Band 9} \label{subsec:Origin_Peak}
One limitation of the multi-wavelength modelling (Section \ref{sec:Multiwave}) is the assumption that the disc is vertically thin and the dust properties (e.g., temperature, grain size) are vertically uniform in the disc. This is a good approximation when high-frequencies observations are optically thin, making them sensitive to the midplane properties. However, if the emission is very optically thick and the disc properties change with height, this assumption might not be valid, as the observations are only sensitive to higher layers of the disc. The effects of vertical temperature gradients or grain size-dependent settling have been explored in \cite{Sierra_2020}.

In this work we assume that all wavelengths are sensitive to the same temperature, and most of their emission and optical depth come from the midplane, which is a good approximation in the optically thin regime. Therefore, the differences in morphology observed in Figure \ref{fig:BrightnessTemperature} should be associated to differences to the dust mass traced at different wavelengths. It is known that different wavelengths are sensitive to different grain sizes  \citep{Draine_2006}. In particular, the DSHARP opacities \citep{Birnstiel_2018} peak at $a_{\rm max} \sim \lambda/ \pi$. Consequently, Bands 9, 7, 6, and 3 in this work are primarily sensitive to grain sizes of 0.14, 0.27, 0.43, and 0.95 mm, respectively. Their Stokes number, assuming a bulk density of 1.675 g cm$^{-3}$ \citep{Birnstiel_2018} are 0.02, 0.05, 0.08, 0.17, respectively, which may exhibit different radial distributions depending on the $\alpha$ parameter \citep[e.g.,][]{Pinilla_2016, Stammler_2023}.

The dust segregation models in Section \ref{sec:DustModels} show that the radial distribution of solids of different sizes peaks at different locations, depending on the disc viscosity. While large grains are trapped in the pressure bump, small grains (constantly generated by fragmentation of large grains) filter through the gap and supply the inner region with dust material. Similar results are obtained from the 2D and 3D hydrodynamical simulations in \cite{Bae_2019} and \cite{Toci_2020}. In the latter, the gas surface density profile drops inward relative to the dust surface density of 1 mm grains. Small grains (not included in their simulation) would be dragged inward and could account for the observed peak shift.
These mechanisms offer a possible explanation for the radial peak offset observed in the Band 9 observations of the disc around PDS 70, and suggest that the PDS 70 b and PDS 70 c may be feeding from these small grains that filter to the disc gap.

Another possible scenario that could explain the peak shift is one where Band 9 observations are more sensitive to temperature gradients than to dust mass. This effect could arise if the Band 9 observations of PDS 70 are optically thick, while observations at longer wavelengths remain optically thin. The inclusion of scattering opacity (discussed in Section \ref{sec:OptDepth}) makes it possible to shift the absorption-only optical depth constraints in Figure \ref{fig:Opt-Depth} into the optically thick regime.

Finally, we cannot rule out the presence of smaller disc substructures that are not traced by the observations in this work and that may shift the disc peak radially inward. As a consistency check, we create dust continuum maps for Bands 7, 6, and 3 at the same angular resolution as Band 9 ($0.242^{\prime\prime}$) and compute their radial profiles. The inner shoulder and the main ring merge into a single ring, and the position of the peak shifts radially inward. However, there remains a discrepancy of approximately 60.5 mas in the peak position between the observations at Bands 7, 6, and 3, compared to that in Band 9. Further observations at higher angular resolution will be necessary to confirm these findings.

\subsection{Origin of the Band 9 outer shoulder}\label{subsec:Shoulder}

The brightness radial profile of PDS 70 (Figure \ref{fig:BrightnessTemperature}) presents an inner and outer shoulder. The former is similar to what is observed at longer wavelengths, although it is also radially offset inward. Its origin is still unclear, but may be tracing a second unresolved ring \citep{Benisty_2021}. The latter is only observed at Band 9, and it is similar to that observed in several discs at high angular resolution \citep[e.g.,][]{Huang_2020, Jennings_2022}. An outer shoulder is also observed in the brightness radial profile of some  optically thin molecules in \cite{Facchini_2021, Rampinelli_2024}, including H$_2$CO, C$_2$H, and DCN, although its radial extent (between $\sim 1.0^{\prime\prime}$ and 1.5$^{\prime\prime}$) is much beyond the shoulder observed at Band 9.

These types of dust continuum shoulders can originate from planet-disc interactions, as shown in the planet-disc simulations of \cite{Bi_2024}, where massive gap-opening planets drive gas radially outward, making coupled grain sizes to escape from the pressure bump by drag forces and form the outer shoulder.
The shoulder is mainly composed of small dust grains, which contribute more significantly to the observed brightness at shorter wavelengths and may explain both the peak shift and the outer shoulder in Band 9.

We remark that both the inner and outer shoulders are only inferred from the visibility modelling when the fit to the power spectrum is high ($w_s \leq 10^{-3}$, Appendix \ref{app:VisFit}). Higher angular resolution observations at this wavelength may be able to confirm or rule out these disc morphologies.

\subsection{Optical depth}
\label{sec:OptDepth}

The multi-wavelength modelling in Section \ref{sec:Multiwave} reveals that the absorption-only optical depth is in the optically thin regime. However, the inclusion of scattering opacity typically increases the total optical depth by at least an order of magnitude, due to the high albedo of millimetre and centimetre sized grains at millimetre wavelengths \citep{Birnstiel_2018}. This shifts the peak values at Bands 9, 7, and 6 into the optically thick regime.

On the other hand, the presence of narrow optically thick regions might be spatially smoothed out in radius. Therefore, the optical depth peaks in Figure \ref{fig:Opt-Depth} should also be considered as lower limits due to beam smearing, particularly in Band 9, where the angular resolution is lowest.
Indirect evidence of optically thick emission in the narrow ring can be observed in the dust continuum maps from Band 3 to 7 (Figure \ref{fig:visibility_modelling}), as the contrast between the azimuthal asymmetry and the ring emission decreases with increasing frequency.

Bright emission along the major axis of discs (as observed in Band 9, Figure \ref{fig:continuum-map}) has been used as observational evidence for optically thin and geometrically thick discs \citep{Doi_2021}. However, the methodology in that work applies only to radially resolved rings, where radial intensity gradients do not significantly contribute to the emission along the major axis, which is primarily biased by the shape of the beam. 
Similar asymmetries along the major axis are observed in Band 6 when imaged at lower angular resolution \citep[Figure 1 in ][]{Facchini_2021}.
Therefore, the asymmetries around the major axis in Band 9 are not direct evidence of optically thin emission at this wavelength.

\section{Conclusions}\label{sec:conclusions}
We analysed new Band 9 observations (671 GHz) of the disc around PDS 70 at an angular resolution of 0.242$^{\prime\prime}$, which in addition to archival infrared and longer millimetre wavelengths observations, provide valuable insights into the size-dependent spatial distribution of solids in the disc. 
Observational constraints, combined with dust evolution models, enhance our understanding of the dust trap mechanism in PDS 70, as well as the leakage of small dust particles into the gap.
The main conclusions obtained from the observational data and dust evolution models are summarized as follows.

\begin{itemize}
    \item The dust continuum map of PDS 70 at Band 9 reveals a ring-like structure and a bright azimuthal asymmetry in the northwest, coinciding with that observed at longer wavelengths.
    Visibility modelling of the disc also reveals an inner shoulder, similar to  observations at longer wavelengths, and an outer shoulder that appears only in the Band 9 observations.
    The known inner disc is neither resolved nor detected at the current resolution and sensitivity.
    
    \item The peak of the Band 9 observations is located at a radius of 0.60$^{\prime \prime}$ (67.4 au), differing from the peak observed at longer millimetre wavelengths (0.68$^{\prime \prime}$,\,76.4 au) and in infrared observations (0.48$^{\prime \prime}$,\,54.0 au). Different wavelengths trace the distribution of various grain sizes in the disk, providing observational evidence of dust segregation and the filtration of small grains through the disk gap.
    
    \item The modelling of the spectral energy distribution at millimetre wavelengths of PDS 70 reveals a dust surface density and maximum grain size peak around the main ring of PDS 70, consistent with dust trap models. The gap is consistent with low dust surface densities and a maximum grain size with large error bars, but consistent with grain sizes ranging from tens of micrometers to centimetres.
    
    \item Dust evolution models of PDS 70 are performed following the model and parameters in \cite{Pinilla_2024}, but for different viscosity parameters $\alpha$. The models show that $\alpha \gtrsim 10^{-3}$ allows small grains to escape from the dust trap and filter through the disc gap. However,  $\alpha \lesssim 5 \times 10^{-3}$ is needed to keep some large grains trapped at the ring location.
    This proof of  concept provides a possible interpretation of why the observed position of the peak emission decreases with wavelength.

    \item 
    The absence of the Band 9 outer shoulder at longer millimetre wavelengths, along with the radially inward offset of the peak, suggests that Band 9 observations of PDS 70 may be primarily sensitive to the spatial distribution of smaller grain sizes (compared to those grains traced at longer wavelengths). These small grains escape from the pressure bump (both inwards and outwards by drag forces), providing observational evidence of a leaky dust trap in the main ring of PDS 70.

\end{itemize}

\section*{Acknowledgments}
We are very thankful for the thoughtful suggestions of the anonymous referee that helped to improve our manuscript significantly.
This paper makes use of the following ALMA data: ADS/JAO.ALMA\#2021.1.01058.S ALMA is a partnership of ESO (representing its member states), NSF (USA) and NINS (Japan), together with NRC (Canada), MOST and ASIAA (Taiwan), and KASI (Republic of Korea), in cooperation with the Republic of Chile. The Joint ALMA Observatory is operated by ESO, AUI/NRAO and NAOJ. The National Radio Astronomy Observatory is a facility of the National Science Foundation operated under cooperative agreement by Associated Universities, Inc.

A.S. and P.P. acknowledge funding from the UK Research and Innovation (UKRI) under the UK government’s Horizon Europe funding guarantee from ERC (under grant agreement No 101076489).
M.B., D.F., and F.Z. have received funding from the European Research Council (ERC) under the European Union's Horizon 2020 research and innovation programme (PROTOPLANETS, grant agreement No. 101002188).
PC acknowledges support by the ANID BASAL project FB210003.
S.F. acknowledges financial contributions by the European Union (ERC, UNVEIL, 101076613), and from PRIN-MUR 2022YP5ACE. Views and opinions are however those of the author(s) only and do not necessarily reflect those of the European Union or the European Research Council. Neither the European Union nor the granting authority can be held responsible for them.
FMe received funding from the European Research Council (ERC) under the European Union's Horizon Europe research and innovation program (grant agreement No. 101053020, project Dust2Planets).
\section*{Software}
This work made use of the following software: CASA \citep{McMullin_2007}, Astropy \citep{astropy:2013, astropy:2018}, Emcee \citep{Foreman_2013}, Frankenstein \citep{Jennings_2020}, Matplotlib \citep{Matplotlib_2007}, Numpy \citep{Numpy_2020}.

\section*{Data availability}
The self-calibrated data underlying this article will be available at \href{http://zenodo.org/}{zenodo.org}. The non-calibrated data is publicly available at \href{http://almascience.org}{almascience.org} using the project code 2021.1.01058.S.



\bibliographystyle{mnras}
\bibliography{main} 



\appendix

\section{More about visibility fit}
\label{app:VisFit}

\begin{figure}
    \centering
    \includegraphics[width=\linewidth]{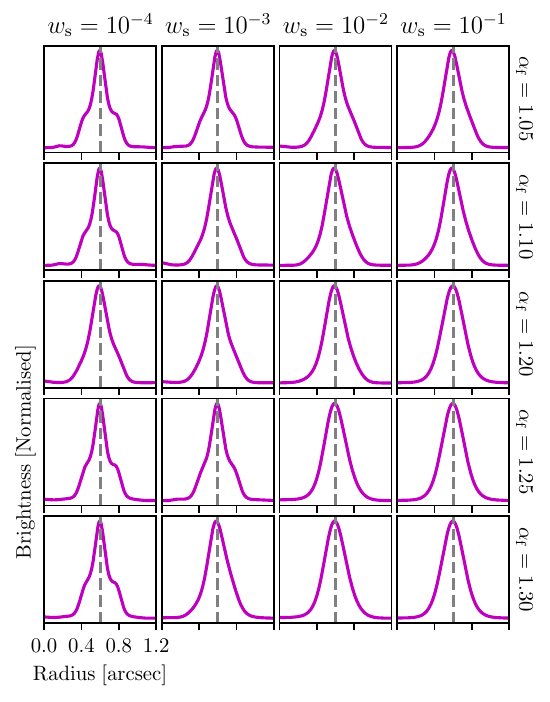}
    \caption{Normalised brightness radial profile obtained from \textsc{Frankenstein} fits for a combination of different hyper-parameters. The dashed vertical dashed line shows the position of the peak at 0.60$^{\prime\prime}$.}
    \label{fig:gallery_vis}
\end{figure}

\textsc{Frankenstein} fits are mainly sensitive to hyper-parameters $\alpha_{\rm f}$ (controlling a threshold to the SNR) and $w_{\rm s}$ (controlling the smoothness of fit to the power spectrum) when the uv-coverage does not include long baselines \citep[$\lesssim 10^{6} \lambda$,][]{Jennings_2020}, as in the case of the Band 9 observations of PDS 70 in this work (Figure \ref{fig:visibility_modelling}). Figure \ref{fig:gallery_vis} shows the reconstructed brightness radial profile for Band 9 using different combination of hyper-parameters.

The position of the peak is at 0.6$^{\prime\prime}$ in all cases. However, the fit is not sensitive to the inner and outer shoulder for high $\alpha_{\rm f}$ or $w_{\rm s}$. The only case where the outer and inner shoulder are only partially observed for $w_{\rm s} = 10^{-4}$ is the one with $\alpha_{\rm f} = 1.2$.
In Figure \ref{fig:BrightnessTemperature} we show the radial profile for the combination of hyper-parameters that provides a good fit to the power spectrum at the longest baselines ($w_{\rm s} = 10^{-4}$) while maintaining a conservative SNR threshold ($\alpha_{\rm f} = 1.3$).

Note that the inner disc is not resolved at this uv-coverage in all cases. Observations covering longer baselines are needed to confirm or rule out the presence of these substructures.

\section{Azimuthal residuals}
\label{app:B9_residuals}

\begin{figure}
    \centering
    \includegraphics[width=\linewidth]{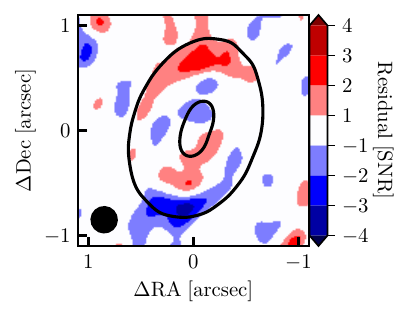}
    \caption{Residual map of PDS 70 at Band 9 after subtracting the axisymmetric model in Figure \ref{fig:BrightnessTemperature}.}
    \label{fig:B9_residual}
\end{figure}

Figure \ref{fig:B9_residual} shows the residual map (Symmetric - Model) of the Band 9 emission, normalised to the rms, also shown in Figure \ref{fig:visibility_modelling}.
All residuals lie within $\pm 4\sigma$, with no significant azimuthal disc structures. 

\section{More about the multi-wavelength fit}
\label{app:DustProperties}
\begin{figure}
    \centering
    \includegraphics[width=0.7\linewidth]{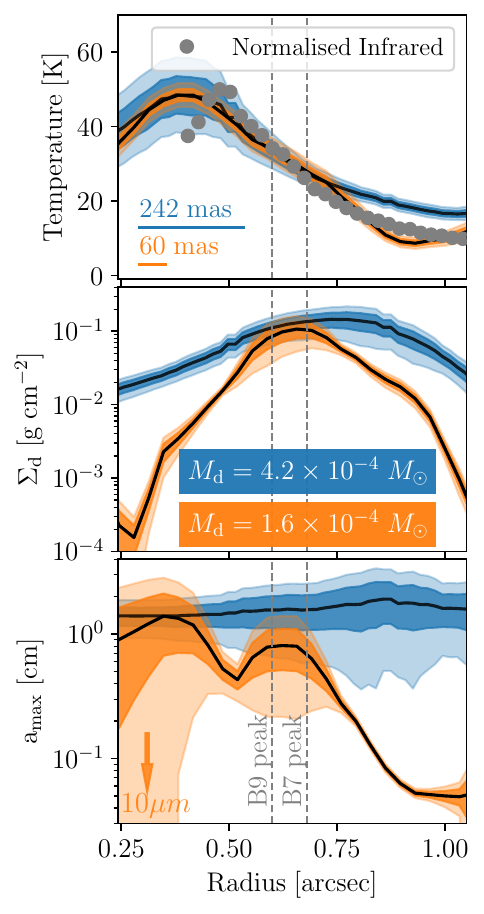}
    \caption{ Dust properties constraints at high angular resolution (orange) and low angular resolution (blue) for the model with $p=3.0$ shown in Figure \ref{fig:Multiwave}. The shaded areas show 1,2 $\sigma$ uncertainties, and the solid line is the best-fit value. The total dust mass is shown in the bottom of the middle panel for each case.}
    \label{fig:DustProperties_2cases}
\end{figure}

Figure \ref{fig:DustProperties_2cases} shows the results from the multi-wavelength fit described in Section \ref{sec:Multiwave}, but for both the low angular resolution fit (blue curves) and high angular resolution fit (orange curves), both in the model with $p=3.0$.
The Band 9 observations are not included in the second fit, but the dust temperature obtained from the former fit is used as a prior distribution for the second fit using Equation \ref{eq:Prior}. This helps to modulate the temperature radial profile when only long-wavelength observations are used, as described in Section \ref{sec:Multiwave}.

\begin{figure}
    \centering
    \includegraphics[width=\linewidth]{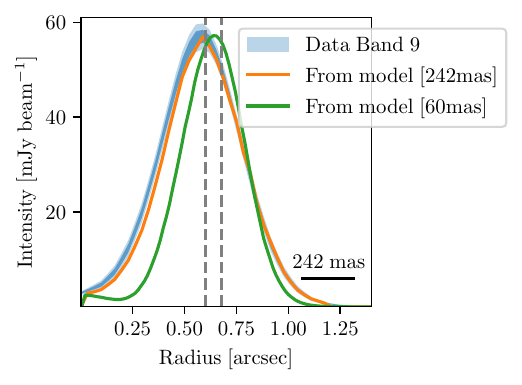}
    \caption{Intensity radial profile in Band 9 derived from the image plane (blue), the best-fit model at $0.242^{\prime\prime}$ resolution (orange), and the best-fit model at $0.060^{\prime\prime}$ resolution (green). The latter has been convolved to match the angular resolution of $0.242^{\prime\prime}$.}
    \label{fig:Predicted_B9}
\end{figure}

Figure \ref{fig:Predicted_B9} shows the observed intensity radial profile in Band 9 and the best-fit models computed from the fits at an angular resolution of $0.242^{\prime\prime}$ and $0.060^{\prime\prime}$ (convolved to the same resolution). The former has a good match with the observed intensity. The latter fails to reproduce the width of the ring and the position of the peak but is able to reproduce the peak's amplitude. Even though the high-angular-resolution fit does not perfectly match the morphology of the Band 9 data, introducing the temperature obtained from the low-angular-resolution fit as a prior helps modulate the flux observed in Band 9.

\bsp	
\label{lastpage}
\end{document}